\documentclass[%
  a4paper, twocolumn, superscriptaddress, prl, amsmath, amssymb, floatfix
]{revtex4-2}
\usepackage{subfiles}
\usepackage{amsthm}
\usepackage{graphicx}
\usepackage{epstopdf}
\usepackage[table]{xcolor}
\usepackage{braket}
\usepackage[colorlinks,bookmarksopen=false]{hyperref}
\usepackage[caption=false]{subfig}
\usepackage{bbm}
\usepackage{makecell}
\usepackage{xr}
\usepackage{siunitx}
\usepackage[normalem]{ulem}
\usepackage{cancel}
\usepackage{times}
\usepackage{booktabs}

\definecolor{darkblue}{HTML}{004D6B}
\definecolor{darkred}{HTML}{8c1515}
\definecolor{darkgreen}{HTML}{006400}

\hypersetup{ colorlinks=true, urlcolor=darkblue, citecolor=darkred,
    linkcolor=darkblue, breaklinks }

\newcommand{\im}{\ensuremath{\textup{i}}}

\newcommand{\rp}{\ensuremath{\mathfrak{Re}}}

\newcommand{\op}[1]{\ensuremath{\mathsf{#1}}}

\newcommand{\pulse}{\mathcal{E}} 

\newtheorem*{mydef*}{Definition}
\newtheorem*{myprop*}{Proposition}

\newtheorem*{mythm*}{Theorem}

\newcommand{\tr}{\ensuremath{\mathrm{tr}}}

\newcommand{\rmin}{\mathrm{in}} 
 \newcommand{\rmc}{\mathrm{c}}
\newcommand{\rmd}{\mathrm{d}}

\begin{document}

\title{%
Chaotic fluctuations in a universal set of transmon qubit gates}

\author{Daniel Basilewitsch}
\affiliation{%
  Dahlem Center for Complex Quantum Systems and Fachbereich Physik, Freie
Universit\"{a}t Berlin, D-14195 Berlin, Germany }

\author{Simon-Dominik B\"orner}
\affiliation{%
  Institute for Theoretical Physics, University of Cologne, D-50937 Cologne,
Germany }

\author{Christoph Berke}
\affiliation{%
  Institute for Theoretical Physics, University of Cologne, D-50937 Cologne,
Germany }

\author{Alexander Altland}
\affiliation{%
  Institute for Theoretical Physics, University of Cologne, D-50937 Cologne,
Germany }

\author{Simon Trebst}
\affiliation{%
  Institute for Theoretical Physics, University of Cologne, D-50937 Cologne,
Germany }

\author{Christiane P. Koch}
\email{christiane.koch@fu-berlin.de}
\affiliation{%
  Dahlem Center for Complex Quantum Systems and Fachbereich Physik, Freie
Universit\"{a}t Berlin, D-14195 Berlin, Germany }

\date{\today}

%TC:ignore 
\begin{abstract}
  Transmon qubits arise from the quantization of nonlinear resonators, systems that are prone to the buildup of  strong, possibly chaotic, fluctuations. Such instabilities will likely affect fast gate operations which involve the transient population of higher excited states outside  the computational subspace. Here we show that a statistical analysis of the instantaneous eigenphases of the time evolution operator, in particular of their curvatures, allows for identifying the subspace affected by chaotic fluctuations.
  Our analysis shows that fast entangling gates, operating at speeds close to the so-called quantum speed limit, contain transient regimes where the dynamics indeed becomes partially chaotic for just two transmons.
\end{abstract}

\maketitle
%TC:endignore 

%%%%%%%%%%%%%%%%%%%%%%%%%%%%%%%%%%%%%%%%%%%%%%%%%%%%%%%%%%%%%%%%%%%%%%%%%%%%%%%%
\paragraph{Introduction}
In coupled superconducting circuits, one of the leading platforms in quantum
information science~\cite{krantz2019quantum, garcia2022quantum}, the nonlinearity of Josephson junctions serves to separate an
energetically low-lying qubit subspace from a higher-lying state space whose
presence has become a focus of attention~\cite{Berke2022,CohenPRXQ2023,Borner2023classical}.
The restriction to the computational subspace
rests on the assumption that the energy stored in the initial state remains
decentralized, and higher excited states of individual `qubits' are not
accessed. However,
residual qubit-qubit couplings can be made algebraically, but never
exponentially, small in the detuning from the resonator frequencies. Indeed, the presence of the higher-lying state space may be detrimental for stationary states relevant to storage of information, in particular so for fixed-frequency architectures~\cite{Berke2022}. While, for information storage, tunable couplers present one route to alleviate signatures of non-integrability, the issue of instabilities due to the non-linearity becomes all the more pressing for the implementation of gates. 

Indeed, the population of energy levels outside of the computational subspace is rather common when implementing fast gates~\cite{StrauchPRL03} or resetting superconducting qubits~\cite{MagnardPRL18}.  Fast
device operation is desirable in view of optimal clock speeds but also to combat
decoherence. The shortest possible time in which a given task can be carried out
is referred to as the quantum speed limit (QSL)~\cite{Deffner2017}. The QSL for 
quantum gates is fundamentally determined by the size of
the effective time-dependent qubit-qubit couplings which
are needed to generate entanglement --- or, in the case of local
rotation gates, to undo undesired entanglement due to residual
couplings. Operation at the QSL is enabled by suitably tailored pulses
which bring the population of higher lying levels during the gate
operation back to the computational subspace at the end of
the gate~\cite{Goerz2017}.
The corresponding time evolution, under the perspective of (classical) nonlinear dynamics, takes, however, place in a twilight zone between integrable dynamics in the limit of weak inter-transmon coupling and possibly chaotic fluctuations at larger couplings~\cite{CohenPRXQ2023,Borner2023classical}. Quantum
mechanically, this ambiguity shows in the structure of transmon many-body
spectra which generically show statistical signatures of both integrability
(evidenced by Poisson statistics) and chaos (Wigner-Dyson
statistics)~\cite{Berke2022}. A transmon gate is then a driven protocol transiently exciting a subsystem of qubits close to the ones targeted by the gate operation. The faster a gate operation, the higher the chances that it connects to semiclassically unstable sectors of the state space~\cite{CohenPRXQ2023}. For gates transiently populating part of the higher-lying state space, instabilities may thus become relevant already in architectures where the computational subspace itself is operated in a stability island~\cite{Berke2022}. While not necessarily reflected in the gate fidelities, the presence of such instabilities will likely impact gate implementation. 
It is thus imperative to clarify the role of the nonlinearity for the dynamics of coupled transmons. 

\begin{figure}[tb]
    \centering
   \includegraphics[width = \columnwidth]{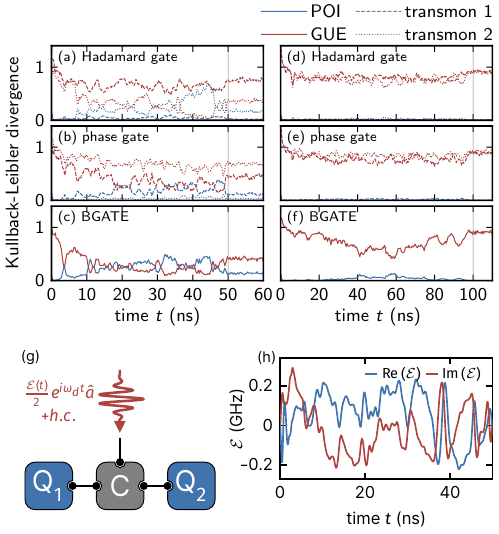}
    \caption{
      \textbf{Spectral analysis of chaotic fluctuations in single-qubit and two-qubit  gates} 
    (a--f) The statistics of the relative eigenphase differences of the time evolution are compared to Poisson (POI), resp.  Gaussian (GUE) statistics using the Kullback-Leibler (KL)
  divergence for gate durations of $T = \SI{50}{\nano\second}$
  (left) and $T = \SI{100}{\nano\second}$ (right). Deviations from Poisson
  statistics are observed in panels (a), (b) and (c) and to a lesser extent
  in panel (f).
  (g) Pictorial view of two transmon qubits
  coupled via a joint transmission line cavity. Gates are implemented by applying a fine-tuned microwave field driving the cavity~\cite{Goerz2017}, shown in (h) for the fast entangling BGATE (c).
  }
  \label{fig:setup+KL}
  \end{figure}

Here, we show how to analyse and detect signatures of non-integrability in a universal set of transmon qubit gates. We adapt methodology from quantum nonlinear dynamics and many-body localization theory to the unitary operators $\op{U}(t)$ implementing single-qubit and two-qubit gates. These time evolutions are realized by time-dependent external fields, designed with optimal control~\cite{Koch2022} as a map on the computational state space, but transiently acting in the larger space of high-lying states. To be specific, we consider the smallest possible structure comprising two transmon qubits coupled via an intermediate cavity subject to a microwave drive~\cite{Goerz2017}, see Fig.~\ref{fig:setup+KL}.
We find that while traces of  nonlinear and even chaotic
dynamics are visible in the operators implementing fast gates,
they present no principal obstruction to high gate fidelities. The reason is
the finite dimensionality of the effective Hilbert space subject to control.
While classical  chaotic time evolution is unpredictable by definition,
projection onto a finite dimensional quantum Hilbert space leads to more manageable dynamics, which can be optimized towards a desired output. Crucially, however,
nonlinearity correlates with a high sensitivity of gate fidelities to just very
small modifications in the Hamiltonian parameters.

%%%%%%%%%%%%%%%%%%%%%%%%%%%%%%%%%%%%%%%%%%%%%%%%%%%%%%%%%%%%%%%%%%%%%%%%%%%%%%%%
\paragraph{Model and control protocols}\label{sec:model}

We analyze two universal sets of gates, with Hadamard and local phase gates
as representative single-qubit operations and the BGATE as
entangling operation. The latter yields the smallest decomposition of an arbitrary two-qubit operation into gates
of a universal set~\cite{ZhangPRL04}. For full operability of
the quantum hardware, it is important to inspect a complete universal set, as
this requires the ability to both create and destroy entanglement.
The Hamiltonian for two transmon qubits
with a joint transmission line cavity subject to a microwave
drive in a frame rotating with the drive frequency
$\omega_{\rmd}$ reads~\cite{Goerz2017}
\begin{eqnarray}
  \op{H}(t) &=& \sum_{q=1}^{2} \left[%
    \delta_{q} \op{b}_{q}^{\dagger} \op{b}_{q}^{\hphantom{\dagger}} +
    \frac{\alpha_{q}}{2} \op{b}_{q}^{\dagger} \op{b}_{q}^{\dagger}
    \op{b}_{q}^{\hphantom{\dagger}} \op{b}_{q}^{\hphantom{\dagger}} + g
    \left(\op{b}_{q}^{\dagger} \op{a} + \op{b}_{q} \op{a}^{\dagger}\right)
    \right] \nonumber \\
  && + \delta_{\mathrm{c}} \op{a}^{\dagger} \op{a} + \frac{1}{2} \pulse(t)
  \op{a} + \frac{1}{2} \pulse^{*}(t) \op{a}^{\dagger}\,,
  \label{eq:H'}
\end{eqnarray}
where $\op{b}_{q}$ and $\op{a}$ are the annihilation operators for the $q$th
transmon and the cavity, respectively, $\delta_{q} = \omega_{q} - \omega_{\rmd}$
and $\delta_{\rmc} = \omega_{\rmc} - \omega_{\rmd}$ are the qubit and cavity
detunings from $\omega_{\rmd}$, $\alpha_{q}$ the qubit
anharmonicities, and $g$ is the qubit-cavity coupling strength. $\pulse(t)$
denotes the amplitude of the cavity drive with the time-dependency derived via gate optimization~\cite{Goerz2017}.
The quantum gates we analyze operate in the quasi-dispersive
straddling qutrits (QuaDiSQ) regime where
the cavity cannot be adiabatically
eliminated and transmon levels outside the computational subspace are exploited for fast gate operation~\cite{Goerz2017} (with dim$\mathcal H=150$, see the supplemental material (SM)~\cite{SM} for details). 
The fasted possible universal set of gates uses gate durations of $T =
\SI{50}{\nano\second}$~\cite{Goerz2017}.  This limit is set by the local gates,
whereas the fastest entangling operation, $\sqrt{i\mathsf{SWAP}}$, requires only
about $\SI{10}{\nano\second}$~\cite{Goerz2017}. Fast gate operations come at
the expense of complex and spectrally broad pulse shapes $\pulse(t)$, see
Fig.~\ref{fig:setup+KL}(h) for an example. Increasing the gate durations by a
factor of two to $T = \SI{100}{\nano\second}$ significantly reduces both
temporal and spectral complexity of the pulses~\cite{Goerz2017}.

%%%%%%%%%%%%%%%%%%%%%%%%%%%%%%%%%%%%%%%%%%%%%%%%%%%%%%%%%%%%%%%%%%%%%%%%%%%%%%%%
\paragraph{Spectral analysis}
\label{subsec:KL}
The time evolution operators $\op{U}(t)$
corresponding to the gate operations
contain information about the entire dynamics up to time $t$, whereas
the instantaneous eigenvalues of the Hamiltonian
~\eqref{eq:H'} capture only time-local information.
We therefore analyze the statistical properties of the instantaneous
eigenvalues of $\op{U}(t)$, using the Kullback-Leibler (KL) divergence~\cite{Kullback1951} to quantify the deviation of their probability distribution from two limiting cases of Poisson (POI) and Gaussian (GUE) statistics for integrable, resp. chaotic, dynamics.
Figure~\ref{fig:setup+KL}(a)--(f) shows the two corresponding KL divergences $D_{\mathrm{KL}}$
(see SM~\cite{SM} for computational details): 
While
$D_{\mathrm{KL}}
(P_{\mathrm{rel.}} | P_{\mathrm{POI}})$ vanishes
almost everywhere for  the slower set of gates (Fig.~\ref{fig:setup+KL}(d)-(f)), the gates at the quantum speed limit (Fig.~\ref{fig:setup+KL}(a)-(c)) show chaotic fluctuations.
\begin{figure*}[tb!]
  \centering
  \includegraphics[scale=0.12]{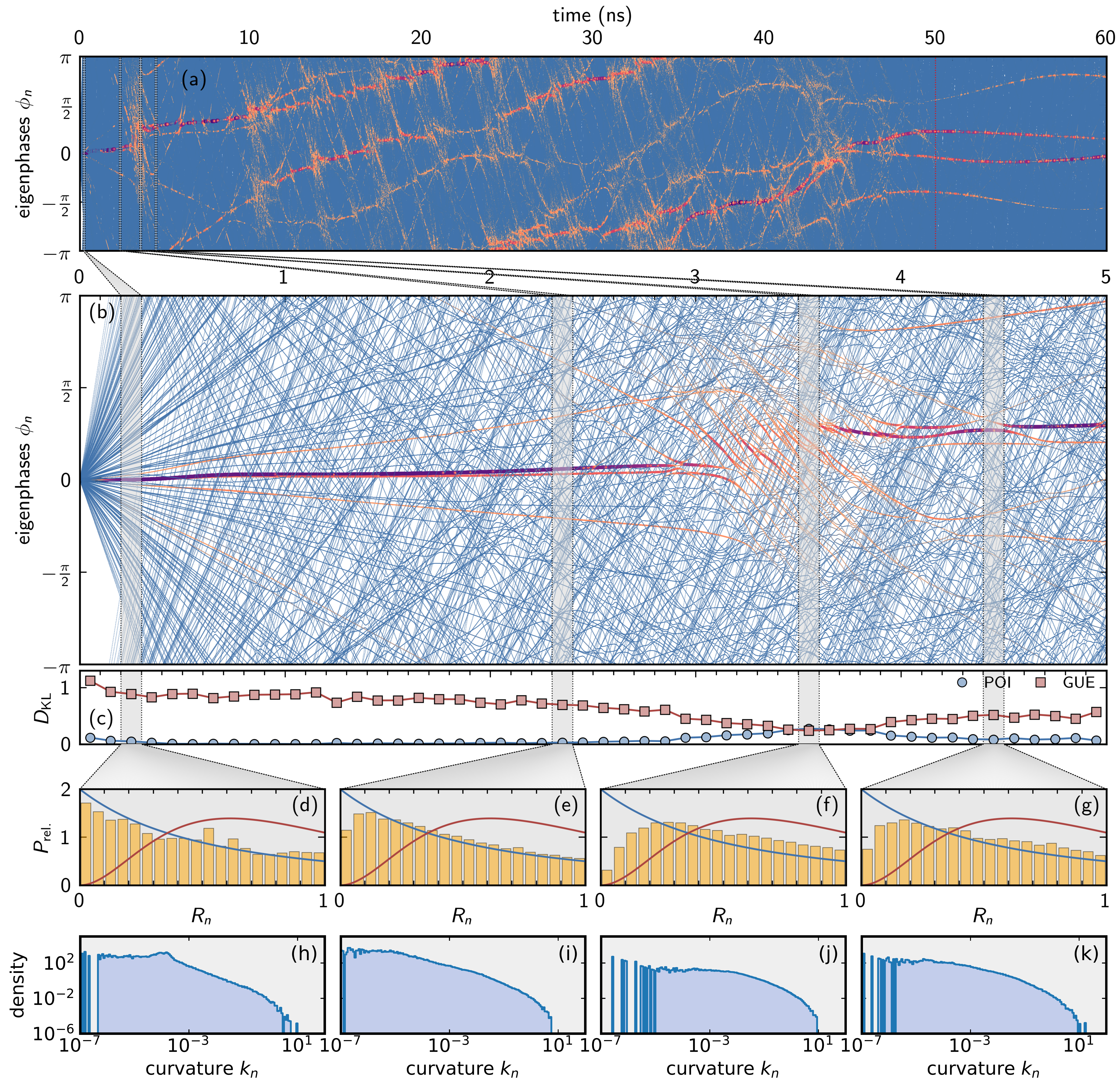}
  \caption{%
    {\bf Detailed analysis of the fast entangling BGATE}
    (c): KL divergences as in
    Fig.~\ref{fig:setup+KL}(c) (red dotted line: end of pulse).
    (a),(b): Instantaneous eigenphases $\{\phi_{n}(t)\}$ of $\op{U}(t)$ (blue lines)
    with their population $p_{i}(t)$ indicated by size and color of the dots.
    (d)--(g): Probability distributions of the relative eigenphases 
    in the time intervals highlighted in
    gray in (b), compared to ideal Poisson (blue) and Gaussian (red) distributions. (h)--(k): Dimensionless level curvatures $k_n(t)$ for the intervals analyzed in (d)--(g). }
  \label{fig:BGATE}
\end{figure*}
Figure~\ref{fig:BGATE}  examines the BGATE, one of the fast gates with chaotic fluctuations, more closely. Panel~
(a) displays the eigenphases $\phi_{n}(t)$ of $\op{U}(t)$ as blue lines, appearing as blue areas  due to their high density, and  the color code showing exemplarily the evolution of the computational basis state $\ket{00}$ (where larger and darker (smaller and brighter) dots indicate a larger (smaller)  overlap 
of the time-evolved state $\ket{\Psi_{00}(t)} = \op{U}(t) \ket{00}$ with the eigenstate $\ket{\Phi_{n}(t)}$ to which the blue line corresponds).
Figure~\ref{fig:BGATE}
(b) provides a blow-up of the first
$\SI{5}{\nano\second}$ of the dynamics. This time frame is
interesting because it shows the transition from straight lines and line
crossings for all $\phi_{n}(t)$ into a regime where avoided crossings between
lines start to become more frequent.
The emerging many-body nonlinear behavior is 
witnessed by the KL divergence of the eigenphases $\phi_{n}(t)$
in Fig.~\ref{fig:BGATE}(c).
While up to $t \approx \SI{3}{\nano\second}$, the KL divergence shows an
almost perfect match with an ideal POI statistics, evidencing integrable
behavior, it diverges from POI statistics and shows a larger match with GUE
statistics, signalling emergent chaotic behavior, between $\SI{3}{\nano\second}$ and
$\SI{4}{\nano\second}$. Figure~\ref{fig:BGATE}
(d)--(g) shows the actual
distributions at four representative times, 
with an almost perfect match with ideal POI statistics before avoided crossings play a role
(d). Similarities of
$P_{\mathrm{rel.}}$ to POI statistics also dominate just before
(e) and after (g) the chaotic regime, whereas  a larger (but far from perfect)  match with GUE statistics 
is observed in
(f). Interestingly, after
$\SI{4}{\nano\second}$ the dynamics becomes less chaotic again. Such emergence
and disappearance of chaos occurs several times over the entire protocol
and is also observed for the other fast gates with chaotic behavior in Fig.~\ref{fig:setup+KL}. Remarkably, the
(re)emergence of chaotic behavior does not have any negative impact on the gate
errors --- these are identically small for both fast and slow protocols
in Fig.~\ref{fig:setup+KL}. The optimized control field seems to navigate the
computational basis states safely through the chaotic regions.

%%%%%%%%%%%%%%%%%%%%%%%%%%%%%%%%%%%%%%%%%%%%%%%%%%%%%%%%%%%%%%%%%%%%%%%%%%%%%%%%
\paragraph{Curvature distributions}
\label{subsec:Curvature}

Beyond the statistics of eigenphases, the structure of avoided crossings can be
another indicator of nonlinear dynamics.
Correlations lead to effective phase repulsion and to an overall more `curvy' pattern of phase evolutions in Fig.~\ref{fig:BGATE}
(b). To quantify this structure, we consider the curvature $\kappa_n$ of the eigenphases,
  $\kappa_n(t) =d^2\phi_n\left(t\right)/dt^2$.
   In the
   limiting cases of localization and chaos,
   they follow distinct distributions~\cite{Filippone2016,Oppen94,Maksymov2019},
   \begin{equation}\label{eq:Curv}
     P_{\mathrm{POI}}(k)=\frac{\mathcal{N}_{\mathrm{POI}}}{\left(1+k^2\right)}\,,\quad
     P_{\mathrm{GUE}}(k)=\frac{\mathcal{N}_{\mathrm{GUE}}}{\left(1+k^2\right)^{2}}\,,
   \end{equation}
   where $\mathcal{N}_{\mathrm{POI/GUE}}$ is a normalization constant and $k$  the curvature $\kappa$, rescaled by the average spacing $\Delta$ and the variance of the velocity distribution
   \cite{Oppen94,Maksymov2019},
   $k_n=\kappa_n \Delta \left( \phi_n\right) /\left(2\pi \mathrm{Var}\left(\partial_t\phi_{n}\right)\right)$.
For large $k$, Eq.~\eqref{eq:Curv} gives a power-law behavior,
 $   P\left(k\rightarrow\infty\right)\propto \left\lVert{k}\right\rVert^{-\widetilde{\beta}}$
with $\widetilde{\beta}=2(4)$ for the Poisson (GUE) case~\cite{GarrattPRB2021,Filippone2016,Maksymov2019}.
\begin{figure}[tb]
  \centering  
   \includegraphics[width=\linewidth]{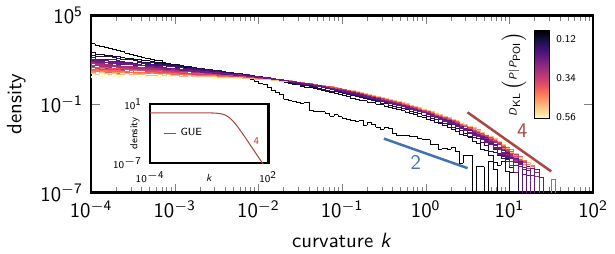}
  \caption{%
    {\bf Curvature statistics} for
    the time evolution implementing the BGATE. The color of the curvature histogram  encodes the KL divergence of the eigenphase spacing. For large curvatures, solid lines indicate polynomial fits to the data, consistent with
   GUE-like behavior (shown in the inset) of a power-law decay with exponent
   $\widetilde{\beta}=4$ for all but the very first time interval which is consistent with
   Poisson-like behavior and $\widetilde{\beta}=2$.}
  \label{fig:curv}
\end{figure}
Analysis of the eigenphase curvatures in Fig.~\ref{fig:BGATE}
(h)--(k) corroborates the occurrence of chaotic fluctuations indicated by the statistical analysis of the eigenphase differences: Noting the double-logarithmic scale, it is clear that small curvatures are prevalent for regular dynamics in Fig.~\ref{fig:BGATE}
(h), (i), and, to a lesser extent (k), whereas larger
curvatures have significantly more weight in the region of nonlinear behavior,
cf. Fig.~\ref{fig:BGATE}(j).
To make the correspondence between the eigenphase statistics and the occurence
of nonlinear/integrable dynamics more quantitative, we calculate the curvature of the eigenphases in each of 30 equally-spaced
$t$ segments of length $\SI{0.5}{\nano\second}$. The resulting distributions of the curvatures are shown
in Fig.~\ref{fig:curv}, where the color code corresponds to the KL divergence
$D(P|P_\text{POI})$ of the eigenphase spacing, calculated for the same time
interval. Importantly, from the perspective of curvatures, there are traces of nonlinear GUE-like
dynamics for all but the initial time intervals. Despite the fact that our
spectrum exhibits only 150 levels, the curvature statistics show general
similarity to pure GUE-behavior (see inset of Fig.~\ref{fig:curv}), exhibiting a
plateau of constant probabilities for small curvatures $k\lesssim 1$ followed by
a crossover to polynomial decay of probabilities in the high curvature regime. While our data does not exhibit a perfectly flat plateau, it
does show the more defining power-law decay for large curvatures, reproducing
the GUE exponent of $\widetilde{\beta}=4$ for all but the initial time interval. The latter
instead follows a POI-like behavior with exponent $\widetilde{\beta}=2$.

We thus find that statistical analysis of the curvatures reflects the presence of avoided crossings in a strikingly clear way. Importantly, it allows for the isolation of a subset of states for which the predictions of random matrix theory apply. Remarkably, this subset of states can comprise only a small portion of Hilbert space. In such cases, the KL divergence has difficulty to indicate the onset of chaotic fluctuations, due to the low weight of the affected states in the sum over all states. This difficulty affects also wave function statistics and out-of-time-ordered correlators which are not sufficiently sensitive as we show in the SM~\cite{SM}.
In contrast, ``sorting" the evolutions according to their curvature allows for identifying the avoided crossings that are a defining hallmark of the chaotic fluctuations. 

%%%%%%%%%%%%%%%%%%%%%%%%%%%%%%%%%%%%%%%%%%%%%%%%%%%%%%%%%%%%%%%%%%%%%%%%%%%%%%%%
\begin{figure}[tb!]
  \centering
  \includegraphics{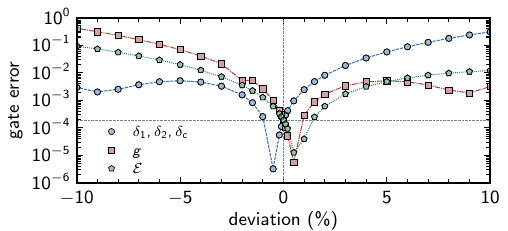}
  \caption{%
    {\bf Gate error for the fast BGATE} with $T = \SI{50}{\nano\second}$ as a
  function of deviations in the Hamiltonian parameters. The horizontal lines
  indicates the gate error in the absence of any deviations. }
  \label{fig:deviations}
\end{figure}
\paragraph{Robustness}
All gate protocols investigated here exhibit remarkably low
gate errors~\cite{Goerz2017}, despite the complexity of  their time-evolution, but the transient and recurrent regimes of instability
identified above are likely to affect gate performance. For ideal protocols, the low errors are readily rationalized from the perspective of quantum optimal control: In a finite-dimensional Hilbert space, it is sufficient to know the exact Hamiltonian that governs the system's dynamics in order to derive high-fidelity control solutions. However, the assumption of precisely knowing
the Hamiltonian is not realistic. For superconducting qubits, in particular,
system parameters are typically not very accurately known and moreover tend to
drift over time. Similar uncertainties apply also to the control
fields, due to inaccuracies in the generating hardware. 
Focussing  on the fast BGATE protocol where  signatures of chaotic behavior have been most apparent, we suspect small deviations in the Hamiltonian or the control parameters to cause large gate
errors. Considering constant deviations,
Fig.~\ref{fig:deviations} analyzes gate robustness with respect to deviations in the transmon and cavity frequencies $\delta_{1}, \delta_{2}, \delta_{\rmc}$, the coupling strength $g$, and the  field amplitude $\pulse(t)$. We 
quantify robustness by the (generalized) gate error,  defined in terms of the local invariants functional~\cite{PhysRevA.84.042315} which measures how much the entangling content of the realized gate deviates from that of the BGATE. In other words, the generalized gate error is insensitive to single-qubit rotations because these are typically easy to correct for.
Despite this relaxed definition,
the BGATE shows high levels of sensitivity in all considered scenarios.
As soon as the deviation exceeds one per cent, the gate error is increased by two to three orders of magnitude, depending on the sign of the deviation (the asymmetry with respect to the sign is readily rationalized by the gate mechanism). 

While the lack of robustness in the fast BGATE may be expectable for a gate showing signatures of chaotic dynamics, it would be premature to conclude that
signatures of chaos in the gate dynamics categorically  hamper gate robustness.
Indeed, the gates of Ref.~\cite{Goerz2017} have not been optimized  to be
robust against parameter fluctuations, and for another two-element model system where chaotic behavior has also been observed, the situation was found to be  controllable \cite{Andreev2021}. 
It will  be interesting to find out under which
conditions robust fast gates can be defined nonetheless, or whether
optimization may identify protocols avoiding the emergence of strong
fluctuations in the spectra and states of $\op{U}(t)$.

%%%%%%%%%%%%%%%%%%%%%%%%%%%%%%%%%%%%%%%%%%%%%%%%%%%%%%%%%%%%%%%%%%%%%%%%%%%%%%%%
\paragraph{Conclusions}
We have found that fast transmon gates
display clear signatures of emergent chaos.
Analysis of the eigenphase curvatures has turned out to be the most sensitive tool to diagnose signatures of non-integrability, due to the ability to identify the subset of states evolving through avoided crossings. More conventional measures, such as the KL divergence of the time evolution eigenphases or measures based on population dynamics, are hampered by an indiscriminate average over state space. For gates that display chaotic fluctuations, time intervals where the statistics have a much larger match with the chaotic limit are followed by time intervals where integrable statistics are recovered. This suggests that the optimized pulses are able to steer the dynamics during a chaotic sea without compromising the gate fidelity. However, when allowing for variations in the parameters characterizing the Hamiltonian, the gate error increases by orders of magnitude already for small deviations. We have thus shown that dynamical instabilities arising from the transmon nonlinearity adversely affect gate implementations already for two qubits and a coupler.

One may wonder whether there is a simple connection between
robustness and signatures of chaos. The answer to this question is less obvious
than one might conjecture at first glance. The quantum speed limit, for
instance, does not impede implementation of energy-efficient quantum
gates~\cite{Aifer2022}. Possibly, energy-efficient quantum gates at the speed
limit would be less prone to chaotic fluctuations than the gates analyzed here. Such gates can be derived by repeating the gate optimization of
Ref.~\cite{Goerz2017}, including energy efficiency as an additional time-dependent constraint~\cite{PalaoPRA13} or explicitly enforcing robustness, e.g. with ensemble optimization~\cite{PhysRevA.90.032329} or by accounting for the parameter fluctuations in the optimization functional~\cite{PhysRevA.104.053118} or equations of motion~\cite{PhysRevA.107.032609}. Generally, robustness of quantum control is key to implementing quantum protocols in real physical systems~\cite{Koch2022,weidner2023robust}. Transmon architectures represent 
a particularly challenging use case, and future work will have to show whether the adverse effects of the transmon nonlinearity can be tamed to harness its benefits for quantum advantage.

%TC:ignore 
%%%%%%%%%%%%%%%%%%%%%%%%%%%%%%%%%%%%%%%%%%%%%%%%%%%%%%%%%%%%%%%%%%%%%%%%%%%%%%%%
\begin{acknowledgments}
  We thank David DiVincenzo for fruitful discussions.
  Financial support from the Deutsche Forschungsgemeinschaft (DFG), 
  Project No.~277101999, CRC 183 (project C05), and the Cluster of
  Excellence Matter and Light for Quantum Computing (ML4Q) EXC 2004/1
  -- 390534769 is gratefully acknowledged. 
\end{acknowledgments}

%%%%%%%%%%%%%%%%%%%%%%%%%%%%%%%%%%%%%%%%%%%%%%%%%%%%%%%%%%%%%%%%%%%%%%%%%%%%%%%%

\section*{Data availability}
The details of the optimized gate protocols and the numerical data shown in the figures are available on Zenodo~\cite{DataZenodo}.

%%%%%%%%%%%%%%%%%%%%%%%%%%%%%%%%%%%%%%%%%%%%%%%%%%%%%%%%%%%%%%%%%%%%%%%%%%%%%%%%

%\bibliography{refs}

%TC:endignore

\newpage
\phantom{(empty page)}
\newpage

\title{%
Supplemental material to ``Chaotic fluctuations in a universal set of transmon qubit gates''}

\author{Daniel Basilewitsch}
\affiliation{%
  Dahlem Center for Complex Quantum Systems and Fachbereich Physik, Freie
Universit\"{a}t Berlin, D-14195 Berlin, Germany }

\author{Simon-Dominik B\"orner}
\affiliation{%
  Institute for Theoretical Physics, University of Cologne, D-50937 Cologne,
Germany }

\author{Christoph Berke}
\affiliation{%
  Institute for Theoretical Physics, University of Cologne, D-50937 Cologne,
Germany }

\author{Alexander Altland}
\affiliation{%
  Institute for Theoretical Physics, University of Cologne, D-50937 Cologne,
Germany }

\author{Simon Trebst}
\affiliation{%
  Institute for Theoretical Physics, University of Cologne, D-50937 Cologne,
Germany }

\author{Christiane P. Koch}
\email{christiane.koch@fu-berlin.de}
\affiliation{%
  Dahlem Center for Complex Quantum Systems and Fachbereich Physik, Freie
Universit\"{a}t Berlin, D-14195 Berlin, Germany }

\date{\today}

\maketitle
\renewcommand{\thefigure}{S\arabic{figure}}
\renewcommand{\theequation}{S\arabic{equation}}

\section{Model and system parameters}

The Hilbert space is constructed as the tensor product of the Hilbert spaces of the two transmons and the cavity. Taking the local Hilbert space dimensions to be 5 for each of the transmons and 6 for the cavity, resulted in a total Hilbert space dimension $N=150$, sufficient for numerical convergence.
The computational states $\ket{00}, \ket{01}, \ket{10}, \ket{11}$ are
the \emph{dressed} basis states, i.e.,
the eigenstates of $\op{H}$ in the absence of the control
which do not change further once the gate
protocol is finished. In detail, $\ket{i_{1} i_{2}}$ refers to the eigenstate
 of $\op{H}$, cf. Eq.~\eqref{eq:H'} in the main text,  with $\pulse(t)=0$, which has
with the largest
overlap with the bare Fock state $\ket{i_{1}} \otimes \ket{i_{2}} \otimes
\ket{i_{\rmc}=0} = \ket{i_{1}, i_{2}, 0}_{\mathrm{Fock}}$, where $i_{1}$,
$i_{2}$ and $i_{\rmc}$ label excitations of the two transmons and the
cavity.

Table~\ref{tab:params} provides an overview of all relevant parameters.
\begin{table}[h!]
  \centering
  \caption{%
    Parameters for the system illustrated in Fig.~\ref{fig:setup+KL}(a) of the main text, consisting of two transmons, $\mathrm{Q}_{1}$ and $\mathrm{Q}_{2}$, and a cavity $\mathrm{C}$. Taken from Ref.~\cite{Goerz2017}.
  }
  \begin{tabular*}{\linewidth}{l@{\extracolsep{\fill}}cr}
    \toprule
    frequency cavity $\mathrm{C}$ & $\omega_{\rmc}/2\pi$ & \SI{6.2}{\giga\hertz}
    \\
    base frequency transmon $\mathrm{Q}_{1}$ & $\omega_{1}/2\pi$ & \SI{6.0}{\giga\hertz}
    \\
    base frequency transmon $\mathrm{Q}_{2}$ & $\omega_{2}/2\pi$ & \SI{5.9}{\giga\hertz}
    \\
    anharmonicity transmon $\mathrm{Q}_{1}$ & $\alpha_{1}/2\pi$ & $-$\SI{290}{\mega\hertz}
    \\
    anharmonicity transmon $\mathrm{Q}_{2}$ & $\alpha_{2}/2\pi$ & $-$\SI{310}{\mega\hertz}
    \\
    coupling between transmons and cavity & $g/2\pi$ & \SI{70}{\mega\hertz}
    \\
    driving frequency of cavity $\mathrm{C}$ & $\omega_{\rmd}/2\pi$ & \SI{5.93}{\giga\hertz}
    \\ \bottomrule
  \end{tabular*}
  \label{tab:params}
\end{table}

%%%%%%%%%%%%%%%%%%%%%%%%%%%%%%%%%%%%%%%%%%%%%%%%%%%%%%%%%%%%%%%%%%%%%%%%%%%%%%%%
\section{Kullback-Leibler divergence of the time evolution eigenphases}\label{sec:KL}

The instantaneous eigenvalues of the time evolution $\op{U}(t)$ corresponding to the gate implementation are denoted by 
$\lambda_{1}(t), \dots,
\lambda_{N}(t) \in \mathbb{C}$,
which have a unique
representation, $\lambda_{n}(t) = e^{\im \phi_{n}(t)}$, $n=1,\dots,N$, in terms
of eigenphases, $\phi_{1}(t), \dots, \phi_{N}(t) \in [-\pi, \pi)$. Assuming without loss of generality  $\phi_{1}(t) \leq \phi_{2}(t) \leq \dots \leq
\phi_{N}(t)$ for every time $t$~\footnote{Note that this sorting might break
continuity if a single eigenphase $\phi_{n}(t)$ is viewed as a function of time}, we define nearest neighbor differences as $\Delta \phi_{n}(t) \equiv
\phi_{n+1}(t) - \phi_{n}(t)$, $n=1,\dots,N-1$. The corresponding relative
differences are then given as~\cite{PhysRevB.75.155111}
\begin{align} \label{eq:Rn}
  r_{n}(t)
  =
  \frac{\Delta \phi_{n}(t)}{\Delta \phi_{n+1}(t)}\,,
\quad
  R_{n}(t)
  =
  \mathrm{min}\left\{r_{n}(t), \frac{1}{r_{n}(t)}\right\}
\end{align}
with $n=1,\dots,N-2$ and $R_{n}(t)\in [0,1]$ which is convenient for
statistical analysis.
In the limiting cases of integrable, resp. chaotic dynamics, the
$\{R_{n}(t)\}$ are expected to obey Poissonian,
respectively Wigner-Dyson statistics. Up to inessential corrections \footnote{Strictly speaking, the distribution of maximally chaotic unitary operators is governed by the circular ensembles~\cite{WeidenmullerRMT}. However, where the statistics of nearby levels is concerned, this point does not matter.}, the corresponding distributions are given
by~\cite{PRL.110.084101}
\begin{subequations} \label{eq:Pideal}
\begin{eqnarray}
  P_{\mathrm{POI}}(R) &=& \frac{2}{(1 + R)^{2}}\,, \\
  P_{\mathrm{GUE}}(R) &=& \frac{162 \sqrt{3}}{4 \pi} \frac{(R + R^{2})^{2}}{(1 +
                          R + R^{2})^{4}}\,.
\end{eqnarray}
\end{subequations}
Deviations from the
limiting cases $P_{\mathrm{POI}}$ or $P_{\mathrm{GUE}}$  can be quantified using
the Kullback-Leibler (KL) divergence, in analogy to the spectral analysis of the
static qubit arrays~\cite{Berke2022}.
More precisely, when comparing to Poisson (POI) statistics, a non-zero KL divergence signals the emergence of statistical correlations. Conversely, chaotic fluctuations are indicated by the KL divergence with respect to $P_\textrm{GUE}$   approaching zero. 
For two statistical distributions $P_{1}$ and $P_{2}$,
their KL divergence is given by~\cite{Kullback1951}
\begin{equation} \label{eq:KL} 
  D_{\mathrm{KL}}(P_{1}|P_{2})
  =
  \sum_{R \in \mathcal{R}} P_{1}(R) \log\left(\frac{P_{1}(R)}{P_{2}(R)}\right)\,,
\end{equation}
where $P_{1}$ and $P_{2}$ are normalized such that $\sum_{R \in \mathcal{R}}
P_{1}(R) = \sum_{R \in \mathcal{R}} P_{2}(R) = 1$. Since in general $D_{\mathrm{KL}}
(P_{1}|P_{2}) \neq D_{\mathrm{KL}} (P_{2}|P_{1})$, we normalize the
KL divergences individually such that $D_{\mathrm{KL}}
(P_{\mathrm{GUE}} | P_{\mathrm{POI}}) = 1$ and $D_{\mathrm{KL}}
(P_{\mathrm{POI}} | P_{\mathrm{GUE}}) = 1$
which ensures $0 \leq
D_{\mathrm{KL}}
(P_{\mathrm{rel.}} | P_{\mathrm{POI}}),
D_{\mathrm{KL}}
(P_{\mathrm{rel.}} | P_{\mathrm{GUE}}) \leq 1$.

To gather enough statistical data, we divide the entire time span into 250 intervals and collect all $R_n$ values from a fixed interval, yielding a single, joint distribution $P_\text{rel.}^\text{joint}$ for each interval. For these joint distributions, the Kullback-Leibler divergences are calculated and displayed in Fig.~\ref{fig:setup+KL}(a--f) of the main text at the respective interval midpoint.

%%%%%%%%%%%%%%%%%%%%%%%%%%%%%%%%%%%%%%%%%%%%%%%%%%%%%%%%%%%%%%%%%%%%%%%%%%%%%%%%
\section{Further observables}\label{sec:further_obs}

For completeness we present here our findings using analysis of wavefunction statistics and out-of-time-ordered correlators. 

%%%%%%%%%%%%%%%%%%%%%%%%%%%%%%%%%%%%%%%%%%%%%%%%%%%%%%%%%%%%%%%%%%%%%%%%%%%%%%%%
\subsection{Occupation dynamics}
\label{subsec:pop}

In order to analyse the dynamics of the states rather than the time-dependent spectra, we introduce and discuss several measures that are based on the population dynamics and show which of those are sensitive regarding the observed non-integrability of the dynamics.

\begin{figure}[b]
  \centering
  \includegraphics{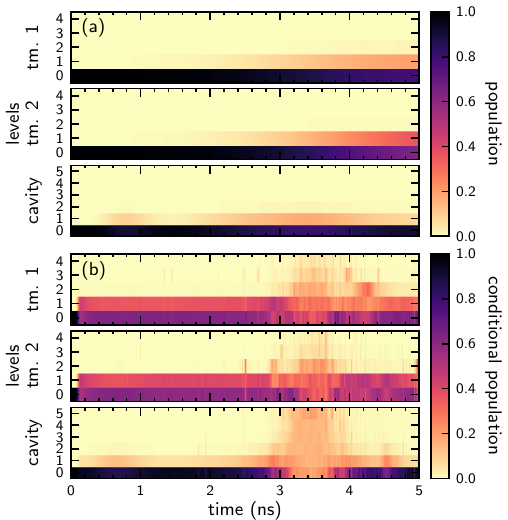}
  \caption{%
    {\bf Occupation dynamics.}
    (a) Occupation dynamics for $\ket{\Psi(t)} = \op{U}(t) \ket{00}$ during the first $\SI{5}{\nano\second}$ of the BGATE. The three panels show the instantaneous, bare level occupations for $\op{\rho}_{1/2/\rmc}(t) = \tr_{2,\rmc/1,\rmc/1,2}\{\ket{\Psi(t)}\bra{\Psi(t)}\}$ of the two transmons (with five levels per transmon) and the cavity (with six levels), respectively.
    (b) The conditional population dynamics, cf. Eq.~\eqref{eq:Pown_eff}, for the same state.
  }
  \label{fig:pop}
\end{figure}

As a first step, we consider the time occupation of the time-evolved
computational basis states $\ket{00}, \ket{01}, \ket{10}, \ket{11}$. The
question is whether traces of the non-integrability we observed in the
time-dependent spectra are visible in the occupation of computational states.
Figure~\ref{fig:pop} (a) shows that this is not the case. Tracing the occupation of a state initialized as $\ket{00}$ over the first $\SI{5}{\nano\second}$ --- an interval which we saw contains regions of regular and non-integrability in the window between $\SI{3}{\nano\second}$ to $\SI{4}{\nano\second}$ ---   we observe no signatures of irregularity in the occupation of states.

The occupation of single particle states close to the ground state thus does not
respond to transient instabilities in the system.
We may therefore rule out the plain occupation dynamics as a good indicator of non-integrability.
However, one may speculate
that the occupation of higher lying states may be a more sensitive
indicator. A first question to be asked is how much of the total spectral weight
carried by an evolving state actually lies outside the logical subspace. To
answer it, we define 
\begin{eqnarray*}
  p_{\mathrm{sub}}(t) &=&
  \frac{1}{4}
  \sum_{\{\ket{\Psi_{\rmin}}\}} \Braket{%
    \Psi_{\rmin} | \op{U}^{\dagger}(t) \op{\Pi}_{\mathrm{sub}} \op{U}(t) |
    \Psi_{\rmin} }\,, \nonumber \\
  \op{\Pi}_{\mathrm{sub}} &=& \sum_{\{\ket{\Psi_{\rmin}}\}} \Ket{\Psi_{\rmin}}
  \Bra{\Psi_{\rmin}}\,,
\end{eqnarray*}
where $\ket{\Psi_{\rmin}} \in \{\ket{00}, \ket{01}, \ket{10}, \ket{11}\}$. Here,
$\op{\Pi}_\textrm{sub}$ is the projector onto the logical subspace and
$p_\textrm{sub}$ the time dependent probability that a state initialized as a
computational state stays inside this space. 

The fast gates analyzed in Fig.~\ref{fig:setup+KL}(a-c) of the main text are characterized for a comparatively large amount of population outside of the logical subspace at intermediate times~\cite{Goerz2017}. For the BGATE, for example, this  figure hovers around 10-20$\%$ at most times, without noticeable changes during the irregular time window. We conclude that the cumulative weight sitting in the non-computational state likewise is blind to dynamical instability. We thus zoom in to the next level of resolution and monitor the occupation of individual ``many-body'' eigenstates $\Phi_n(t)$ of the full evolution operator $\op{U}(t)$.  With an initial state as before, $\ket{\Psi_{\rmin}} \in \{\ket{00}, \ket{01}, \ket{10}, \ket{11}\}$, we define the probabilities   
\begin{align} \label{eq:p_inst}
  p_{n}(t)
  =
  \left|\Braket{\Phi_{n}(t) | \op{U}(t) | \Psi_{\rmin}}\right|^{2},
\end{align} 
normalized as $\sum_n p_n=1$. The color coding in Fig.~\ref{fig:BGATE}(a) and
(b) of the main text shows the distribution of these weights over the eigenphases $\phi_n$ of the
first 150 states of the system. In this representation, blue color is
used to indicate the bare eigenphases, while a combined color coding from dark
purple to light orange and dot size coding from larger to smaller dots indicates
larger to smaller state occupation.
This visual data indeed shows a massive fragmentation of the evolving state over a high
dimensional subspace during intervals where the spectral analysis flags
non-integrability (via the KL divergence). Outside these regions, the spectral
weight remains concentrated on a few states in and outside the computational
subspace. However,  while the correlation of state fragmentation and
non-integrability is a generic phenomenon, we have also observed milder forms of
transient state spreading in gate protocols without chaotic instabilities, such
as the slower operations shown in Fig.~\ref{fig:setup+KL}(d)--(f) of the main text.

While the above data may serve as a visual indicator for dynamical signatures of non-integrability, it contains too much information to be quantitatively useful. The simplest way to condense it is to track the time dependent fraction $M /N$ of the number of levels,  $M$, whose  occupation probability exceeds a given threshold $p$: 
\begin{align}
  \label{eq:Pcount}
  P_{p}(t)=\frac{1}{N}\sum_{n=1}^N \Theta(p_n(t)-p), 
\end{align}   
where $\Theta$ is the Heaviside step function. Taking $p=1\%$,  $P_p(t)$ reflects the spreading/contracting of the colored dots in Fig.~\ref{fig:BGATE}(a) of the main text. Inspecting this figure is, however, also 
relatively non-expressive in that it is qualitatively similar to the total occupation of the out-of-computational subspace $1-P_\textrm{sub}$, hinting at a limited usefulness of Eq.~\eqref{eq:Pcount}.

A more meaningful quantity is obtained by  the projection of exact
instantaneous eigenstates $\Phi_n$  onto select occupation eigenstates,
$\ket{i_{1}, i_{2}, i_{\rmc}}_{\mathrm{Fock}}$, weighted with the occupation
probabilities $p_n$ computed according to Eq.~\eqref{eq:p_inst},
\begin{align} \label{eq:Pown}
  P_{i_{1}, i_{2}, i_{\rmc}}(t)
  &=
  \sum_{n}
  p_{n}(t)
  \left|\Braket{\Phi_{n}(t) | i_{1}, i_{2}, i_{\rmc}}_{\mathrm{Fock}}\right|^{2} \,,
\end{align}
Note that Eq.~\eqref{eq:Pown}
is quartic in the states $\{\ket{\Phi_{n}(t)}\}$. (A squared amplitude is hiding in $p_n$.) 
We read these fourth moments of wave functions as a sum over conditional
probabilities, where the first factor $p_n$ states the representation of state
$\Phi_n$ in the actual time evolved initial state, and the second its
probability to be found in the Fock space state $\ket{i_{1}, i_{2},
i_{\rmc}}_{\mathrm{Fock}}$.  These quantities can be further reduced to yield,
say, the representation of the first transmon's states as 
\begin{align} \label{eq:Pown_eff}
  P_{i_{1}}(t)
  =
  \sum_{i_{2}=0}^{N_{2}-1}
  \sum_{i_{\rmc}=0}^{N_{\rmc}-1}
  P_{i_{1}, i_{2}, i_{\rmc}}(t) \,,
\end{align}
Fig.~\ref{fig:pop} (b) shows how this data efficiently detects the instability
in the high lying sector of Hilbert space during the interval between 3 and
4~ns. In this way, it represents a good compromise between
the bare occupation probabilities of the low lying states, which are blind to
the presence of unstable regimes (Fig.~\ref{fig:pop} (b)), and the excessive
data carried by the full set of amplitudes $\{\Phi_n\}$. 

To summarize, we have found two state population-based measures, Eqs.~\eqref{eq:p_inst} and~\eqref{eq:Pown_eff}, that are sensitive to the non-integrability of the dynamics. Unfortunately, both of them require the diagonalization of $\op{U}(t)$ and their computation will therefore become difficult for larger systems. In contrast, the analyses based solely upon the occupation dynamics, i.e., both level or subspace occupations, show no correlation with the observed irregularities. This suggests that such irregularities might be easily overseen if only the occupation dynamics is analyzed.

\section{Out-of-time-ordered correlators (OTOCs)}
\label{app:otocs}

\begin{figure}[tb]
  \centering
  \includegraphics{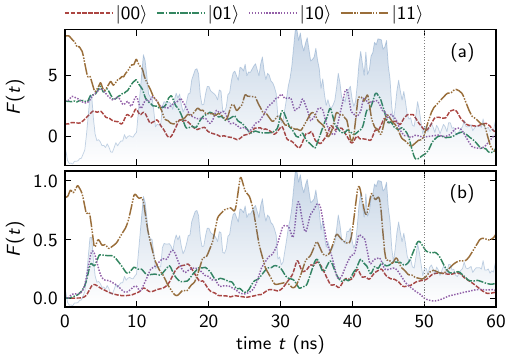}
  \caption{%
    {\bf OTOCs for the fast BGATE} with $T = \SI{50}{\nano\second}$. The
    operators are (a) $\op{V} = \op{b}_{1} + \op{b}_{1}^{\dagger}$ and $\op{W} =
    \op{b}_{2} + \op{b}_{2}^{\dagger}$ as well as (b) $\op{V} =
    \op{b}_{1}^{\dagger} \op{b}_{1}$ and $\op{W} = \op{b}_{2}^{\dagger}
    \op{b}_{2}$. The blue area in the background indicates the KL divergence for
    the POI statistics from Fig.~\ref{fig:setup+KL} (e) of the main text in arbitrary units. }
  \label{fig:otocs}
\end{figure}
Additionally, we check so-called out-of-time-ordered correlators (OTOCs), which
are a good measure for information scrambling as well as quantum
chaos~\cite{Swingle2018}. Let $\op{V}$ and $\op{W}$ be two Hermitian operators,
the OTOC is given by
\begin{align} \label{eq:otoc}
  F(t)
  =
  \rp\left\{
    \Braket{%
      \op{W}^{\dagger}(t) \op{V}^{\dagger} \op{W}(t) \op{V}
    }_{\Psi}
  \right\},
  \quad
  \op{W}(t)
  =
  \op{U}^{\dagger}(t) \op{W} \op{U}(t),
\end{align}
which depends on the state $\ket{\Psi}$ for which it is evaluated. If $\op{V}$
and $\op{W}$ are also unitary, besides being Hermitian, and fulfill $[\op{V},
\op{W}] = 0$, we have $F(0) = 1$ for all $\ket{\Psi}$ and a deviation like $F(t)
< 1$ at later times $t$ indicates information scrambling between the two
subspaces acted upon by $\op{V}$ and $\op{W}$. While in two-level systems a
possible choice would be $\op{V}, \op{W} \in \{\op{\sigma}_{x}, \op{\sigma}_{y},
\op{\sigma}_{z}\}$ such that $\op{V}$ and $\op{W}$ are both Hermitian \emph{and}
unitary. Unfortunately, this is note possible in our case due to the larger
Hilbert space. However, since unitarity of $\op{V}$ and $\op{W}$ is not a
requirement, we choose
\begin{subequations}
\begin{equation}
  \op{V}
  =
  \op{b}_{1} + \op{b}_{1}^{\dagger},
  \qquad
  \op{W}
  =
  \op{b}_{2} + \op{b}_{2}^{\dagger}
\end{equation}
and
\begin{equation}
  \op{V}
  =
  \op{b}_{1}^{\dagger} \op{b}_{1},
  \qquad
  \op{W}
  =
  \op{b}_{2}^{\dagger} \op{b}_{2}.
\end{equation}
\end{subequations}
Figure~\ref{fig:otocs} shows the OTOCs for those two choices of $\op{W}$ and
$\op{V}$ for all four computational basis states $\ket{00}, \ket{01}, \ket{10},
\ket{11}$ for the fast BGATE. The blue background area indicates the KL
divergence with respect to the POI statistics in order to compare the course of
the OTOCs to peaks in the KL divergence indicating chaotic regions. We find the
OTOCs to not match the KL divergence, neither for the two choices of $\op{V}$
and $\op{W}$ presented in Fig.~\ref{fig:otocs} nor for other choices.

%\bibliography{refs}

\bibliography{refs}
\end{document}